\documentstyle[12pt]{article}
\date{}

\newtheorem{theorem}{Theorem}

\newcommand{\lam}{{\lambda}}
\newcommand{\blam}{{\bar\lambda}}
\newcommand{\De}{{D'}}

\newcommand{\vps}{{\varepsilon}}
\newcommand{\bvps}{{\bar\varepsilon}}
\newcommand{\bA}{{\bar A}}
\newcommand{\al}{\alpha}
\newcommand{\bal}{{\bar\alpha}}
\newcommand{\be}{\beta}
\newcommand{\bbe}{{\bar\beta}}
\newcommand{\de}{\delta}
\newcommand{\bde}{{\bar\delta}}
\newcommand{\bxi}{{\bar\xi}}

\newcommand{\bm}{{\bar m}}

\newcommand{\sig}{{\sigma}}
\newcommand{\bga}{{\bar\gamma}}
\newcommand{\bsig}{{\bar\sigma}}
\newcommand{\btau}{{\bar\tau}}
\newcommand{\bome}{{\bar\omega}}
\newcommand{\bmu}{{\bar\mu}}
\newcommand{\bnu}{{\bar\nu}}
\newcommand{\bpi}{{\bar\pi}}
\newcommand{\bka}{{\bar\kappa}}
\newcommand{\brho}{{\bar\rho}}
\newcommand{\eth}{{\not\!\partial}}
\newcommand{\beth}{{\bar{\not\!\partial}}}
\newcommand{\bPsi}{{\bar\Psi}}
\newcommand{\tPsi}{{\tilde\Psi}}
\newcommand{\bze}{{\bar\zeta}}
\newcommand{\sY}{{{\ }_sY}}
\newcommand{\ppr}{{\frac{\partial}{\partial r}}}
\newcommand{\ppu}{{\frac{\partial}{\partial u}}}
\newcommand{\ppz}{{\frac{\partial}{\partial\zeta}}}
\newcommand{\ppbz}{{\frac{\partial}{\partial{\bar\zeta}}}}
\newcommand{\bee}{\begin{eqnarray}}
\newcommand{\ede}{\end{eqnarray}}
\begin{document}
\baselineskip=14 pt

\begin{center}
{\Large\bf On Uniqueness of Kerr Space-time near null infinity}
\end{center}
\begin{center}
Xiaoning Wu${}^a$\footnote{e-mail : wuxn@amss.ac.cn} and Shan
Bai${}^b$\footnote{e-mail : xiaobai99033@yahoo.com.cn}
\end{center}
\begin{center}
a. Institute of Mathematics,\\ Academy of Mathematics and Systems
Science,\\ Chinese Academy of Sciences,\\ P.O.Box 2734, Beijing,
100080, China.\\
{\ }\\
b. Institute of Theoretical Physics, \\
Chinese Academy of Sciences,\\ P.O.Box 2735, Beijing,
100080, China.\\
\end{center}
\begin{abstract}
We re-express the Kerr metric in standard Bondi-Saches' coordinate
near null infinity ${\cal I}^+$. Using the uniqueness result of
characteristic initial value problem, we prove the Kerr metric is
the only asymptotic flat, stationary, axial symmetric, Type-D
solution of vacuum Einstein equation. The Taylor series of Kerr
space-time is expressed in terms of B-S coordinates and the N-P
constants have been calculated.
\end{abstract}

{\bf keywords}: Kerr solution, Uniqueness Theorem



\section{Introduction}
After the work by Bondi et.al.\cite{Bon}, it is well-known the Bondi
coordinates is a very natural choice when we want to describe the
asymptotic behavior of gravitational field near null infinity ${\cal
I}^+$. Based on works by Penrose, Newman and Unti\cite{PR86,UN62},
there is an elegant way to re-express Bondi's work in N-P formulism.
This also gives us a general formulism to describe the asymptotic
structure of general asymptotic flat space-time which is smooth
enough near ${\cal I}^+$. Using characteristic initial value (CIV)
problem method, many authors\cite{Fr81,Ka96,Ni06} have shown the
existence of null infinity in general case and pointed out the
degree of freedom of gravitational field near null infinity. The CIV
method has many advantages in dealing with gravitational radiation
problem. Recently, this method has been used in numerical
relativity. Winicour and his colleagues develop the CCM method. They
want to combine the CIV method and standard Hamiltonian evolution
method\cite{Wi05}. On the other hand, Kerr solution is a very
important exact solution of Einstein equation both in theoretical
area and in application. An interesting question is whether Kerr
metric describes the space-time outside a stationary rotating star.
For long time, the Bondi coordinates of Kerr space-time is not very
clear. For example, how to describe the Kerr space-time in
Unti-Newman's general formalism\cite{UN62}? The uniqueness
theorem\cite{Ro75,He96} tells us that Kerr solution is the only
asymptotic flat, stationary, axial symmetric solution of vacuum
Einstein equation with regular event horizon. From the application
point of view, it is very difficult to get the detail information
about the event horizon of a space-time because of the infinite
red-shift near horizon. An interesting question is how to identify
the Kerr solution based on information near null infinity. This is
more practical in future gravitational experiments. This idea can
also be understood from the Geroch conjecture\cite{Ger70}.
Obviously, stationary and axial symmetric condition is not enough
because there are many asymptotic flat exact solutions of Ernst
equation. In next section, it is found such uncertainty comes from
the homogeneous part of a control equation which comes from the
Killing equation. The general solutions of that equation contain
some free constants. These unknown constants are close related with
Geroch-Hansen multi-pole moments\cite{Fr06}. In order to identify
the Kerr solution, we use Petrov classification\cite{Kr80} and show
that condition will help us to pick out Kerr solution finally.
Further more, N-P constants for Kerr space-time is also calculated
as a byproduct.

This paper is organized as following : In section II, we prove an
local uniqueness theorem of Kerr solution based on the information
near null infinity. This theorem also tells us the standard
Unti-Newman expansion of Kerr metric. The detail expression of this
extension is contained in Appendix A. Appendix B contains some
spin-weight harmonics which is useful for our calculation.

\section{Main theorem}
Let $(M,g)$ be an asymptotic flat space-time, $(u,r,\theta,\varphi)$
be the standard Bondi-Sachs coordinates.

\begin{theorem}
Suppose $(M,g)$ be an asymptotic flat, stationary, axial symmetric,
Type-D, vacuum space-time with in a neighborhood of null infinity,
then it is isometric to Kerr space-time in the Bondi coordinates
neighborhood.
\end{theorem}

In order to make the proof clearly enough, we divide it into two
subsections. In first subsection, we calculate the Taylor series of
general stationary axially symmetric space-time in Bondi
coordinates. We find all Taylor coefficients can be expressed in
terms of $\{\Psi^k_0\}$ and their derivatives. Unknown functions
$\{\Psi^k_0\}$ satisfy a linear inhomogeneous second order partial
differential equation. We find the general solution of such equation
has the form $\Psi^{k}_0=\tPsi^{k}_0+D^k{\ }_2Y_{k+2,0}$, where
$\tPsi^k_0$ is the special solution which corresponds to Kerr metric
and $D^k$ is a free constant, i.e. all axially symmetric and
stationary solutions are characterized by the set of constant
$\{D^k\}$. This set of constants are close related with the famous
Geroch-Hansen multipole. In section 2.2, with the help of Type-D
condition, we show all $\{D^k\}$ vanish and Kerr metric is the
unique stationary, axial-symmetric, asymptotic flat Tpye-D metric .

\subsection{Taylor series of general axial symmetric vacuum stationary spacetime}

Suppose $(M,g)$ be a vacuum stationary axial symmetric space-time.
$t^a$ and $\phi^a$ are two commutative Killing vectors. Near ${\cal
I}^+$, we use the standard B-S coordinates to do the standard
asymptotic expansion. The detailed construction of these coordinates
is well-known and can be found in Re.\cite{PR86,UN62}. With this
choice of coordinates, we also can choose a set of null tetrad
$\{l^a,n^a,m^a,\bm^a\}$, such that
$l^a=\left(\frac{\partial}{\partial r}\right)^a$ and these tetrad
are parallel-transported along $l^a$. Under such choice of
coordinates,
$\phi^a=\left(\frac{\partial}{\partial\varphi}\right)^a$. The
time-like Killing vector $t^a$ can be expressed in terms of null
tetrad as $t^a=Tl^a+Rn^a+A\bm^a+\bA m^a$. $[t^a,\phi^a]=0$ means
that $T,R,A$ are independent of $\varphi$. The null tetrad
components of Killing equation for $t^a$ are
\begin{eqnarray}
&&-DR+(\vps+\bvps)R+\bka A+\kappa\bA=0,\nonumber\\
&&-DT-(\vps+\bvps)T-\pi A-\bpi\bA-\De R+(\gamma+\bga)R+\btau
A+\tau\bA=0,\nonumber\\
&&-\kappa T+\bpi R+DA+(\bvps-\vps)A-\de R+(\bal+\be)R+\brho
A+\sig\bA=0,\nonumber\\
&&-\De T-(\gamma+\bga)T-\nu A-\bnu\bA=0,\nonumber\\
&&-\tau T+\bnu R+\De A+(\bga-\gamma)A-\de T-(\bal+\be)T-\mu
A-\blam\bA=0,\nonumber\\
&&-\sig T+\blam R+\de A+(\bal-\be)A=0,\nonumber\\
&&-\rho T+\mu R+\de\bA-(\bal-\be)\bA-\brho T+\bmu R+\bde
A-(\al-\bbe)A=0.\nonumber
\end{eqnarray}
Using the Bondi Gauge $\kappa=\vps=\pi=0$, $\rho=\brho$,
$\tau=\bal+\be$, we have
\begin{eqnarray}
&&-DR=0,\label{a1}\\
&&-DT-\De R+(\gamma+\bga)R+\btau
A+\tau\bA=0,\label{a2}\\
&&DA-\de R+\tau R+\brho
A+\sig\bA=0,\label{a3}\\
&&-\De T-(\gamma+\bga)T-\nu A-\bnu\bA=0,\label{a4}\\
&&-\tau T+\bnu R+\De A+(\bga-\gamma)A-\de T-\tau T-\mu
A-\blam\bA=0,\label{a5}\\
&&-\sig T+\blam R+\de A+(\bal-\be)A=0,\label{a6}\\
&&-\rho T+\mu R+\de\bA-(\bal-\be)\bA-\brho T+\bmu R+\bde
A-(\al-\bbe)A=0.\label{a7}
\end{eqnarray}
Here we use the standard notation of \cite{PR86,UN62,Kr80}.
Differential operators in above equations are defined as
\begin{eqnarray}
D&:=&\frac{\partial}{\partial r},\nonumber\\
\De&:=&\frac{\partial}{\partial u}+U\frac{\partial}{\partial
r}+X\frac{\partial}{\partial \zeta}+{\bar
X}\frac{\partial}{\partial\bze},\nonumber\\
\de&:=&\omega\frac{\partial}{\partial
r}+\xi^3\frac{\partial}{\partial\zeta}+\xi^4\frac{\partial}{\partial\bze},\quad
 \zeta=e^{i\varphi}\cot\frac{\theta}{2}.
\end{eqnarray}

It is well known that stationary solutions to Einstein's vacuum
field equations are analytic\cite{Ha70}. Moreover, it is also known
that asymptotically flat stationary vacuum solutions are not only
analytic, but even admit an analytic conformal extension through
null infinity\cite{DS90,Dain01}. Keeping this result in mind, all
geometric quantities (the coordinate components of null tetrad, N-P
coefficients, components of time-like Killing vector and components
of Weyl curvature) can be expressed in terms of power series of
$\frac{1}{r}$, for example
\begin{eqnarray}
T&=&T^0+\frac{T^1}{r}+\cdots,\nonumber\\
A&=&A^0+\frac{A^1}{r}+\cdots\ ,\label{TA}
\end{eqnarray}
some lower order Taylor coefficients of components of tetrad, N-P
coefficients and components of Weyl tensor (up to 3th order) can be
found in section 9.8 of \cite{PR86}.

First of all, let's consider the function $R$. Eq.(\ref{a1}) and
axial symmetric condition tell us that $R=R(u,\theta)$. With the
formal expansion of null tetrad and N-P coefficients\cite{PR86,
UN62}, zero order of Eq.(\ref{a2}) gives
\begin{eqnarray}
\frac{\partial R}{\partial u}=0,
\end{eqnarray}
so $R=R(\theta,\phi)$. In order to get more information about $R$,
higher order of Killing equation are needed. The first order of
Eq.(\ref{a3}), Eq.(\ref{a5}) and Eq.(\ref{a7}) are
\begin{eqnarray}
\de_0R+A^0&=&0,\label{11}\\
-\bPsi^0_3R+{\dot
A}^1-\de_0T^0+\frac{1}{2}A^0-{\dot\sig}^0\bA^0&=&0,\label{12}\\
2T^0-R&=&0,\label{13}
\end{eqnarray}
where
$\de_0=\frac{(1+\zeta\bze)}{\sqrt{2}}\frac{\partial}{\partial\bze}$
 and `` $\cdot$ " means ${\partial}_u$
. Because the space-time is stationary, there is no Bondi flux. This
implies ${\dot\sig}^0=0$ \cite{PR86}, then N-P equations tell us
this leads
 $\Psi^0_3=0$. Combining this condition with
(\ref{11}),(\ref{12}) and (\ref{13}), we get
\begin{eqnarray}
{\dot A}^1+A^0=0.
\end{eqnarray}
The second order of Eq.(\ref{a3}) is
\begin{eqnarray}
2A^1=\frac{\zeta(1+\zeta\bze)}{\sqrt{2}\ \bze}\ \sig^0\frac{\partial
R}{\partial\zeta}+\sig^0\bA^0.
\end{eqnarray}
It has been shown that the right hand side of above equation is
independent on $u$, so ${\dot A}^1=0$. This implies $A^0=0$ and $R$
is a constant. So we can take $R=1$ without lost of generality.

With $R=1$, Killing equations become
\begin{eqnarray}
&&-DT+(\gamma+\bga)+\btau
A+\tau\bA=0,\label{2}\\
&&DA+\tau+\brho
A+\sig\bA=0,\label{3}\\
&&-\De T-(\gamma+\bga)T-\nu A-\bnu\bA=0,\label{4}\\
&&-\tau T+\bnu+\De A+(\bga-\gamma)A-\de T-\tau T-\mu
A-\blam\bA=0,\label{5}\\
&&-\sig T+\blam+\de A+(\bal-\be)A=0,\label{6}\\
&&-\rho T+\mu+\de\bA-(\bal-\be)\bA-\brho T+\bmu+\bde
A-(\al-\bbe)A=0.\label{7}
\end{eqnarray}
With Eq.(\ref{TA}), the non-trivial zero order Killing equation is
\begin{eqnarray}
{\dot T}^0=0.
\end{eqnarray}

Non-trivial first order Killing equations are
\begin{eqnarray}
&&-A^0=0,\\
&&-{\dot T}^1+\Psi^0_3A^0+\bPsi^0_3\bA^0=0,\\
&&-\bPsi^0_3+{\dot A}^1+\frac{1}{2}A^0-{\dot\bsig}^0A^0=0,\\
&&{\dot\bsig}^0+\de_0A^0+2\bal^0A^0=0,\\
&&2T^0-1=0,
\end{eqnarray}
which implies ${\dot T}^1=0$ and ${\dot A}^1=0$. Same order N-P
equations also give $\Psi^0_3=0$ and $\Psi^0_4=0$.

Second order Killing equations are
\begin{eqnarray}
&&T^1-\frac{1}{2}(\Psi^0_2+\bPsi^0_2)=0,\\
&&-2A^1=0,\\
&&{\dot T}^2=0,\\
&&\frac{1}{2}\eth\Psi^0_2+{\dot A}^2-\de_0T^1=0,\\
&&\frac{1}{2}\sig^0=0,\\
&&2T^1-\sig^0{\dot\bsig}^0-\bsig^0{\dot\sig}^0-\Psi_2^0-\bPsi^0_2=0,
\end{eqnarray}
where $\eth$ is the spin-weight operator\cite{PR86} and is defined
as $\eth f:=(\de_0+2s\bal^0)f$,
$\al^0=-\frac{\cot\theta}{2\sqrt{2}}$ . From these equations, we
know $\sig^0=0$, $A^1=0$, ${\dot T}^2=0$,
$T^1=\frac{1}{2}(\Psi^0_2+\bPsi^0_2)$, $\Psi^0_2=\bPsi^0_2$,
${\dot\Psi}^0_2=0$, ${\dot
A}^2=-\frac{1}{2}\eth\Psi^0_2+\frac{1}{2}\de_0(\Psi^0_2+\bPsi^0_2)$.
It is worth to point out that the result ${\dot\sig}^0=0$ tells us
that the Bondi coordinates which is chosen as \cite{UN62} are
associated with the ``good cut" of stationary spacetime, i.e.
$\sig^0=0$, so the freedom of super-translation has been removed.

The third order killing equations are
\begin{eqnarray}
&&2T^2+\frac{1}{3}(\beth\Psi^0_1+\eth\bPsi^0_1)=0,\label{23}\\
&&-3A^2-\frac{1}{2}\Psi^0_1=0,\label{33}\\
&&{\dot T}^3=0,\label{43}\\
&&\frac{1}{2}\Psi^0_1+\bnu^3+{\dot
A}^3+\frac{3}{2}A^2-\de_0T^2=0,\label{53}\\
&&\de_0A^2+2\bal^0A^2=0,\label{63}\\
&&2T^2+\frac{1}{2}\beth\Psi^0_1+\frac{1}{2}\eth\bPsi^0_1+\de_0\bA^2-2\bal^0\bA^2+\bde_0A^2-2\al^0A^2=0.
\end{eqnarray}
Eq.(\ref{33}),(\ref{63}) imply
\begin{eqnarray}
\eth\Psi^0_1=0.\label{Psi1}
\end{eqnarray}
The spin-weight of $\Psi^0_1$ is $1$, so it is a linear combination
of spin-weight harmonics $\{{}_1Y_{l,m}\}$. The axial symmetric
condition implies $m=0$. The behavior of spin-weight harmonic under
the action of operators $\eth$ and $\beth$ are \cite{PR86}
\begin{eqnarray}
\eth\sY_{lm}&=&-\sqrt{\frac{(l+s+1)(l-s)}{2}}{\ }_{s+1}Y_{lm},\nonumber\\
\beth\sY_{lm}&=&\ \ \sqrt{\frac{(l-s+1)(l+s)}{2}}{\
}_{s-1}Y_{lm},\nonumber\\
{\ }_0Y_{lm}&=&Y_{lm}.\label{swh}
\end{eqnarray}
So we get $\Psi^0_1=c(u){}_1Y_{1,0}=c(u)\sin\theta$. Detailed
calculation on same order N-P equations also give
\begin{eqnarray}\nu^3=-\frac{1}{12}\bPsi^0_1-\frac{1}{6}\beth^2\Psi^0_1,\
\Psi^1_3=0,\ \Psi^2_3=\frac{1}{2}\beth^2\Psi^0_1, \
\Psi^1_4=\Psi^2_4=\Psi^3_4=0,\
\Psi^4_4=-\frac{1}{4}\beth\Psi^3_3.\label{psi23psi44}
\end{eqnarray}
Eq.(\ref{23}) and ${\dot T}^2=0$ gives
\begin{eqnarray}
{\dot c}\beth\sin\theta+{\dot{\bar c}}\eth\sin\theta=2({\dot
c}+{\dot{\bar c}})\cos\theta=0,
\end{eqnarray}
which implies ${\dot c}+{\dot{\bar c}}=0$. The first order Bianchi
identities tell us ${\dot\Psi}^0_1-\eth\Psi^0_2=0$, which implies
$\Psi^0_2=-{\dot c}\cos\theta+C_2$ but it is well-known that
${\bar\Psi}^0_2=\Psi^0_2$ for stationary case\cite{PR86,UN62}, so
${\dot c}=0$. This means $\Psi^0_1=C_1\sin\theta$ and
$\Psi^0_2=C_2$. The Komar integral shows $-C_2$ is just the Bondi
mass of the space-time. Additionally, Eq.(\ref{53}) tells us that
${\dot A}^3=0$.

Forth order Killing equations are
\begin{eqnarray}
&&3T^3+(\gamma^4+\bga^4)=0,\label{24}\\
&&4A^3=\frac{1}{3}\beth\Psi^0_0,\label{34}\\
&&{\dot T}^4+\frac{1}{3}(\beth\Psi^0_1+\eth\bPsi^0_1)(\Psi^0_2+\bPsi^0_2)=0,\label{44}\\
&&\frac{1}{2}\Psi^0_1T^1-\frac{1}{3}\beth\Psi^0_0+\bnu^4+{\dot
A}^4+(\Psi^0_2+\bPsi^0_2)A^2+\frac{3}{2}A^3-\de_0 T^3+\Psi^0_2A^2+\frac{1}{2}A^3=0,\label{54}\\
&&\frac{1}{4}\Psi^0_0+\blam^4+\eth A^3=0,\label{64}\\
&&2T^3+\mu^4+\bmu^4+\eth\bA^3+\beth A^3=0.\label{74}
\end{eqnarray}
where
$\gamma^4=-\frac{1}{12}(\al^0\beth\Psi^0_0-\bal^0\eth\bPsi^0_0)-\frac{1}{8}\beth^2\Psi^0_0$,
$\lam^4=-\frac{1}{12}\bPsi^0_0$,
$\mu^4=-\frac{1}{3}\Psi^2_2=-\frac{1}{6}\beth^2\Psi^0_0$,
$\nu^4=\frac{1}{24}(\eth\bPsi^0_0+\beth^3\Psi^0_0)$,
$\Psi^3_3=-\frac{1}{2}\bPsi^0_1\Psi^0_2-\frac{1}{6}\beth^3\Psi^0_0$
(These results are got from same order N-P equations).

The spin-weight of $\Psi_0$ is 2, so Eq.(\ref{34}),(\ref{64}) imply
\begin{eqnarray}
\Psi^0_0=D^5(u)\sin^2\theta,\label{Psi00}
\end{eqnarray}
Eq.(\ref{53}),(\ref{34}) eliminate time dependence of $D^5(u)$, i.e.
$\Psi^0_0=D^5\sin^2\theta$. Eq.(\ref{54}) is \bee {\dot A}^4=0 \ede

Fifth order Killing equations are \bee
&&4T^4+(\gamma^5+\bga^5)-\frac{1}{2}\bPsi^0_1A^2-\frac{1}{2}\Psi^0_1\bA^2=0,\label{25}\\
&&A^4=\frac{1}{5}\tau^5,\label{35}\\
&&-\frac{1}{2}\sig^5+\blam^5+\frac{1}{2}\Psi^0_0T^1+\frac{3}{2}\Psi^0_1A^2+\eth A^4=0,\label{65}\\
&&-2\rho^5+2T^4+(\mu^5+\bmu^5)+\frac{3}{2}\Psi^0_1\bA^2+\frac{3}{2}\bPsi^0_1A^2
+\eth\bA^4+\beth A^4=0,\label{75}
\ede
where
\bee \rho^5&=&0,\nonumber\\
\mu^5&=&-\frac{1}{4}\Psi^3_2,\nonumber\\
\sig^5&=&-\frac{1}{3}\Psi^1_0,\nonumber\\
\lam^5&=&-\frac{1}{8}\bPsi^0_0\Psi^0_2-\frac{1}{24}\bPsi^1_0,\nonumber\\
\gamma^5&=&-\frac{1}{40}(\al^0\beth\Psi^1_0-\bal^0\eth\bPsi^1_0)+\frac{1}{12}|\Psi^0_1|^2-\frac{1}{30}\beth^2\Psi^1_0,
\nonumber\\
\tau^5&=&\frac{1}{8}\beth\Psi^1_0,\nonumber\\
\Psi^3_2&=&-\frac{2}{3}|\Psi^0_1|^2+\frac{1}{6}\beth^2\Psi^1_0,\nonumber\\
\Psi^2_1&=&-\frac{1}{2}\beth\Psi^1_0.\nonumber \ede Eq.(\ref{25}),
(\ref{44}) and Bianchi identities imply $\Psi^0_1=iC_1\sin\theta,\
C_1\in{\bf R}$, where $C_1$ is the Komar angular momentum.
Eq.(\ref{35}),(\ref{65}) give
\begin{eqnarray}
\eth\beth\Psi^1_0+5\Psi^1_0=10(\Psi^0_1)^2-15\Psi^0_0\Psi^0_2.\label{Psi01}
\end{eqnarray}
The homogeneous part of above equation is
\begin{eqnarray}
\eth\beth\Psi^1_0+5\Psi^1_0=0.
\end{eqnarray}
Because spin-weight of $\Psi^1_0$ is $2$, it is a linear combination
of $\{{}_2Y_{l,0}\}$. Using Eq.(\ref{swh}), the homogeneous equation
is
\begin{eqnarray}
(-l^2-l+12)\ {}_2Y_{l,0}=0,
\end{eqnarray}
which gives $l=3$. The general solution of Eq.(\ref{65}) is
\begin{eqnarray}
\Psi^1_0=\left(\frac{10}{3}(C_1)^2-5C_2D^5\right)\sin^2\theta+D^6\
{}_2Y_{3,0}.
\end{eqnarray}
 (Bianchi identities insures
${\dot\Psi}^1_0=0$.) By definition\cite{PR86}, the non-zero N-P
constant for stationary axial symmetric space-time is
\begin{eqnarray}
G_0=\frac{10}{3}(C_1)^2-5C_2D^5.\label{npc}
\end{eqnarray}

Until now, we have got series expression of tetrad components up to
4th order, N-P coefficients up to 5th order and Weyl components up
to 6th order. To prove this theorem, all Taylor coefficients of all
geometric quantities are needed. We use inductive method to solve
this problem order by order.

Suppose we have known Taylor coefficients of tetrad components up to
$(k-3)^{th}$ order, Taylor coefficients of connections up to
$(k-2)^{th}$ order and Taylor coefficients of Weyl curvature
components up to $(k-1)^{th}$ order. The $(k-1)^{th}$ order of
Killing equation (\ref{3}) and (\ref{6}) are
\begin{eqnarray}
-(k-1)A^{k-2}+\tau^{k-1}=\cdots,\label{ac1}\\
\cdots+\blam^{k-1}+\eth A^{k-2}=0.\label{ac2}
\end{eqnarray}
where ``$\cdots$" means terms which only contain lower order
coefficients. Based on the induction hypothesis, those terms are
known. In order to solve these equations, we need coefficients
$\lam^{k-1}$ and $\tau^{k-1}$. From N-P equations,
\begin{eqnarray}
D\Psi_1-\bde\Psi_0=-4\al\Psi_0+4\rho\Psi_1&\Rightarrow&\Psi^{k}_1=-\frac{1}{(k-4)}\beth\Psi^{k}_0+\cdots.\nonumber\\
D\sig=2\rho\sig+\Psi_0&\Rightarrow&\sig^{k-1}=-\frac{1}{(k-3)}\Psi^{k}_0+\cdots,\nonumber\\
D\lam=\rho\lam+\bsig\mu&\Rightarrow&\lam^{k-1}=\frac{1}{2(k-2)}\ \bsig^{k-1}+\cdots,\nonumber\\
D\tau=\tau\rho+\btau\sig+\Psi_1&\Rightarrow&\tau^{k-1}=-\frac{1}{(k-2)}\Psi^{k}_1+\cdots,
\label{coek}
\end{eqnarray}
Combining Eq.(\ref{ac1}), (\ref{ac2}) and (\ref{coek}), we get
\begin{eqnarray}
\eth\beth\Psi^{k}_0+\frac{(k+4)(k+1)}{2}\Psi^{k}_0=\cdots.\label{Psik0}
\end{eqnarray}
The homogeneous part of above equation is
\begin{eqnarray}
\eth\beth{\hat\Psi}^{k}_0+\frac{(k+4)(k+1)}{2}{\hat\Psi}^{k}_0=0.
\end{eqnarray}
Because of Eq.(\ref{swh}) and axial symmetric condition, the general
solution should be
\begin{eqnarray}
\Psi^{k}_0=\tPsi^{k}_0+D^k{\ }_2Y_{k+2,0}\ ,\label{GPsi0k}
\end{eqnarray}
where $\tPsi^k$ is a special solution of eq.(\ref{Psik0}) and $D^k$
is a constant. Obviously, Kerr solution satisfies all conditions of
our theorem, so $\tPsi^k_0$ must exist. The concrete form of
$\tPsi^k_0$ also can be got by direct calculation. One can express
the ``$\cdots$" terms in Eq.(\ref{Psik0}) as a linear combination of
spin-weight harmonics $\{{}_2Y_{l,0}\}$. The inductive method
insures the maximal value of $l$ in that expression will be finite
for any given order, then we can get $\tPsi^k_0$ by comparing
coefficients between bother sides of this equation. With the general
solution of $\Psi^k_0$, Eq.(\ref{coek}) will give $\tau^{k-1}$,
$\sig^{k-1}$, $\lam^{k-1}$ and $\Psi^k_1$. Further more, Cartan
structure equations and Bianchi equations will help us to get other
coefficients,
\begin{eqnarray}
D\rho=\rho^2+|\sig|^2&\Rightarrow&-(k-3)\rho^{k-1}=\cdots,\nonumber\\
D\al=\al\rho+\be\bsig&\Rightarrow&-(k-2)\al^{k-1}=\cdots,\nonumber\\
D\be=\be\rho+\al\sig+\Psi_1&\Rightarrow&-(k-2)\be^{k-1}=\Psi^k_1+\cdots,\nonumber\\
D\Psi_2-\bde\Psi_1=3\rho\Psi_2-2\al\Psi_1-\lam\Psi_0&\Rightarrow&-(k-3)\Psi^k_2=\beth\Psi^k_1+\cdots,\nonumber\\
D\Psi_3-\bde\Psi_2=2\rho\Psi_3-2\lam\Psi_1&\Rightarrow&-(k-2)\Psi^k_3=\beth\Psi^k_2+\cdots,\nonumber\\
D\Psi_4-\bde\Psi_3=\rho\Psi_4+2\al\Psi_3-3\lam\Psi_2&\Rightarrow&-(k-1)\Psi^k_4=\beth\Psi^k_3+\cdots,\nonumber\\
D\gamma=\tau\al+\btau\be+\Psi_2&\Rightarrow&-(k-1)\gamma^{k-1}=\Psi^k_2+\al_0\tau^{k-1}-\bal_0\btau^{k-1}+\cdots,\nonumber\\
D\mu=\mu\rho+\lam\sig+\Psi_2&\Rightarrow&-(k-2)\mu^{k-1}=\frac{1}{2}\rho^{k-1}+\Psi^k_2+\cdots,\nonumber\\
D\nu=\tau\lam+\btau\mu+\Psi_3&\Rightarrow&-(k-1)\nu^{k-1}=\frac{1}{2}\btau^{k-1}+\Psi^k_3+\cdots,\nonumber\\
D\xi^3=\rho\xi^3+\sig\bxi^4&\Rightarrow&-(k-3)\xi^3_{k-2}=\cdots,\nonumber\\
D\xi^4=\rho\xi^4+\sig\bxi^3&\Rightarrow&-(k-3)\xi^4_{k-2}=\cdots,\nonumber\\
D\omega=\rho\omega+\sig\omega-(\bal+\be)&\Rightarrow&-(k-3)\omega^{k-2}=-\bal^{k-1}-\be^{k-1}+\cdots,\nonumber\\
DX=(\bal+\be)\xi^3+(\al+\bbe)\bxi^4&\Rightarrow&-(k-2)X^{k-2}=\frac{1+|\zeta|^2}{\sqrt{2}}(\al^{k-1}+\bbe^{k-1})+\cdots,\nonumber\\
DU=(\bal+\be)\bome+(\al+\bbe)\omega-\gamma-\bga&\Rightarrow&-(k-2)U^{k-2}=-\gamma^{k-1}-\bga^{k-1}+\cdots.\label{Gk}
\end{eqnarray}
From above results, we find we can express all $(k-2)^{th}$ order
coefficients of tetrad components, $(k-1)^{th}$ order coefficients
of connection components and $k^{th}$ order coefficients of Weyl
curvature in terms of $\Psi^k_0$, $\eth$ derivatives of $\Psi^k_0$
and lower order coefficients which we have known. The form of
$\Psi^k_0$ is given in Eq.(\ref{GPsi0k}). This means we can get all
those coefficients for any given order.

\subsection{Uniqueness of Kerr solution}

In above subsection, we have got Taylor series of a general
stationary axial symmetric metric. From Eq.(\ref{GPsi0k}), we can
see that the freedom in each order Taylor coefficients are just the
constant $D^k$ ($\ k\ge 5$). These arbitrary constants should be
closely related to the famous Geroch-Hansen multi-pole
moments\cite{Ger70,Han74,Ku88,Fr06}. What we want to do in this
section is to pick out the Kerr solution from those possible
solutions, i.e. we need to fix value of $\{D^k\}$. In order to do
that, we consider the Petrov classification\cite{Kr80}. It is well
known that the Kerr solution belongs to Type-D class, i.e. its Weyl
curvature satisfies\cite{Kr80}
\begin{eqnarray}
K=\Psi_1(\Psi_4)^2-3\Psi_4\Psi_3\Psi_2+2(\Psi_3)^3=0.\label{typed}
\end{eqnarray}
Write down the $12^{th}$ order coefficient of above equation, we get
\begin{eqnarray}
-3\Psi^4_4\Psi^2_3\Psi^0_2+2(\Psi^2_3)^3=0.\label{typed0}
\end{eqnarray}
In previous section, we have got
\begin{eqnarray}
&&\Psi^0_2=C_2,\nonumber\\
&&\Psi^0_1=iC_1{\ }_1Y_{1,0},\nonumber\\
&&\Psi^2_3=\frac{1}{2}\beth^2\Psi^0_1,\nonumber\\
&&\Psi^0_0=D^5{\ }_2Y_{2,0},\nonumber\\
&&\Psi^3_3=-\frac{1}{2}\bPsi^0_1\Psi^0_2-\frac{1}{6}\beth^3\Psi^0_0,\nonumber\\
&&\Psi^4_4=-\frac{1}{4}\beth\Psi^3_3.\label{a4}
\end{eqnarray}
This constant is fixed in following way : from Komar integrals
$M=\frac{1}{8\pi}\int_{S_{\infty}}*dt$ and
$J=Ma=\frac{1}{16\pi}\int_{S_{\infty}}*d\phi$, we know
$\Psi^0_2=-M$, $\Psi^0_1=3iMa\sqrt{\frac{4\pi}{3}}{\ }_1Y_{1,0}$.
Submit these into Eq.(\ref{a4}) then get
\begin{eqnarray}
&&\Psi^2_3=\frac{3iMa}{2}\sqrt{\frac{4\pi}{3}}{\
}_{-1}Y_{1,0}\quad,\nonumber\\
&&\Psi^3_3=\left(\frac{3iM^2a}{2}\sqrt{\frac{4\pi}{3}}-\frac{1}{\sqrt{2}}D^5\right){\ }_{-1}Y_{2,0}\quad,\nonumber\\
&&\Psi^4_4=\left(-\frac{i\sqrt{6\pi}M^2a}{4}+\frac{D^5}{4}\right){\
}_{-2}Y_{2,0}\quad.
\end{eqnarray}
Submit above result into Eq.(\ref{typed0}), we find
\begin{eqnarray}
0&=&3M\left(-\frac{i\sqrt{6\pi}M^2a}{4}+\frac{D^5}{4}\right){\
}_{-2}Y_{2,0}+2\left[\frac{3iMa}{2}\sqrt{\frac{4\pi}{3}}{\
}_{-1}Y_{1,0}\right]^2\nonumber\\
&=&\left[3M\left(-\frac{i\sqrt{6\pi}M^2a}{4}+\frac{D^5}{4}\right)-\frac{3M^2a^2}{4}\sqrt{\frac{16\pi}{5}}\right]{\
}_{-2}Y_{2,0}
\end{eqnarray}
Solving the simple linear algebraic equation
$\left[3M\left(-\frac{i\sqrt{6\pi}M^2a}{4}+\frac{D^5}{4}\right)-\frac{3M^2a^2}{4}\sqrt{\frac{16\pi}{5}}\right]=0$,
we can fix the value of $D^5$ is
\begin{eqnarray}
D^5={Ma^2}\sqrt{\frac{16\pi}{5}}+{i\sqrt{6\pi}M^2a}.
\end{eqnarray}
Submit above result into Eq.(\ref{npc}), it is easy to check the N-P
constant of Kerr space-time is zero, which has been got by
\cite{WS06, Bai06}.

In order to fix general $D^k$, we also use inductive method and fix
them order by order. Suppose we have known $D^k$ up to order $n$. To
get the value of $D^{n+1}$, we consider the $(n+8)^{th}$ coefficient
of Eq.(\ref{typed}), a long but direct calculation shows it should
be
\begin{eqnarray}
-3\Psi^{n+1}_4\Psi^2_3\Psi^0_2+\cdots=0,\label{n+1}
\end{eqnarray}
here ``$\cdots$" also mean terms which only contain lower order
coefficients. From Eq.(\ref{coek}),(\ref{GPsi0k}),(\ref{Gk}), we
know
\begin{eqnarray}
&&\Psi^{n+1}_0=\tPsi^{n+1}_0+D^{n+1}{\ }_2Y_{n+3,0},\nonumber\\
&&\Psi^{n+1}_1=-\frac{1}{(n-3)}\beth\Psi^{n+1}_0+\cdots.\nonumber\\
&&\Psi^{n+1}_2=-\frac{1}{n-2}\beth\Psi^{n+1}_1+\cdots,\nonumber\\
&&\Psi^{n+1}_3=-\frac{1}{n-1}\beth\Psi^{n+1}_2+\cdots,\nonumber\\
&&\Psi^{n+1}_4=-\frac{1}{n}\beth\Psi^{n+1}_3+\cdots,\label{last}
\end{eqnarray}
where $\tPsi^{n+1}_0$ is the special solution of Eq.(\ref{Psik0})
which corresponds to Kerr solution. We have known Kerr solution
belongs to Type-D, i.e. Eq.(\ref{n+1}) holds for $\tPsi^{k}_0$, so
Eq.(\ref{n+1}) can be written as
\begin{eqnarray}
\frac{-3\Psi^2_3\Psi^0_2}{(n-3)(n-2)(n-1)n}\beth^4\tPsi^{n+1}_0+\cdots=0.
\end{eqnarray}
If the general $\Psi^{n+1}_0$ in Eq.(\ref{last}) also satisfies
Eq.(\ref{n+1}), i.e.
\begin{eqnarray}
\frac{-3\Psi^2_3\Psi^0_2}{(n-3)(n-2)(n-1)n}\beth^4\tPsi^{n+1}_0+D^{n+1}\frac{-3\Psi^2_3\Psi^0_2}{(n-3)(n-2)(n-1)n}\beth^4{\
}_2Y_{n+3,0}+\cdots=0.
\end{eqnarray}
Because terms in ``$\cdots$" only contain lower order coefficients,
they remain unchanged when we change $\tPsi^{n+1}_0$ to the general
$\Psi^{n+1}_0$. Obviously,
$\frac{-3\Psi^2_3\Psi^0_2}{(n-3)(n-2)(n-1)n}\beth^4{\ }_2Y_{n+3,0}$
is a non-zero function for any $n$, so the general solution of
$\Psi^{n+1}_0$ satisfies Eq.(\ref{n+1}) means $D^{n+1}=0$ and
$\Psi^{n+1}_0=\tPsi^{n+1}_0$. This tells us that Kerr solution is
the only solution which satisfies all requirements of our theorem.

Remark : in above subsection, we proved our theorem under the
requirement that the whole space-time is type-D. In subsection 2.2,
we have shown that the freedom of vacuum, stationary,
axial-symmetric space-time are just a set of constants $\{D^k\}$.
The reason why we need the condition type-D is to fix the value of
$\{D^k\}$. If the type-D condition holds at several points in Bondi
neighborhood and those points are not zero-points of
$\frac{-3\Psi^2_3\Psi^0_2}{(n-3)(n-2)(n-1)n}\beth^4{\ }_2Y_{n+3,0}$,
then, it is easy to see that $\{D^k\}$ should be zero, i.e. type-D
condition holds at several points will imply this condition holds in
the whole neighborhood. This feature may be a practical method from
the experiment prospective. That means only several points need to
be checked for the type-D condition to see whether the space-time
around us is Kerr space-time. This maybe a useful property for
future gravitational experiments, such as ``mapping
space-time"\cite{Bab06}.

\section*{Acknowledgement}
This work is supported by the Natural Science Foundation of China
under Grant Nos.10705048, 10605006, 10731080 and K.C.Wong Education
Foundation, Hong Kong. Authors would like to thank Prof. X.Zhang and
Dr. J.A.Valiente-Kroon for their helpful discussion.

\section*{Appendix A}
The asymptotic extension of Kerr space-time in B-S coordinates are

1) Null tetrad
\begin{eqnarray}
l^a&=&\ppr,\nonumber\\
n^a&=&\ppu+\left[-\frac{1}{2}+\frac{M}{r}-\frac{Ma^2}{2r^3}(3\cos^2\theta-1)+O(r^{-4})\right]\ppr\nonumber\\
&&+\left[\frac{iMa}{2r^3}\cot\frac{\theta}{2}+O(r^{-4})\right]\ppz
+\left[-\frac{iMa}{2r^3}\cot\frac{\theta}{2}+O(r^{-4})\right]\ppbz,\\
m^a&=&\left[-\frac{3iMa}{2\sqrt{2}r^2}\sin\theta+\frac{Ma^2}{\sqrt{2}r^3}\sin^2\theta+O(r^{-4})\right]\ppr\nonumber\\
&&+O(r^{-4})\frac{\partial}{\partial\zeta}
+\left[\frac{(1+\zeta\bze)}{\sqrt{2}r}+O(r^{-4})\right]\frac{\partial}{\partial\bze}.\nonumber
\end{eqnarray}

2) N-P coefficients
\begin{eqnarray}
\rho&=&-\frac{1}{r}+O(r^{-5}),\nonumber\\
\sig&=&-\frac{3Ma^2\sin\theta}{2r^4}+O(r^{-5}),\nonumber\\
\al&=&-\frac{\cot\theta}{2\sqrt{2}r}+\frac{3Ma^2\sin\theta\cos\theta}{2\sqrt{2}r^4}+O(r^{-5}),\nonumber\\
\beta&=&\frac{\cot\theta}{2\sqrt{2}r}-\frac{3iMa\sin\theta}{2\sqrt{2}r^3}+\frac{33Ma^2\sin\theta\cos\theta}{2\sqrt{2}r^4}
+O(r^{-5}),\nonumber\\
\tau&=&-\frac{3iMa\sin\theta}{2\sqrt{2}r^3}+\frac{18Ma^2\sin\theta\cos\theta}{\sqrt{2}r^4}
+O(r^{-5}),\nonumber\\
\lam&=&-\frac{Ma^2\sin^2\theta}{4r^4}+O(r^{-5}),\nonumber\\
\mu&=&-\frac{1}{2r}+\frac{M}{r^2}+\frac{3iMa\cos\theta}{2r^3}-\frac{Ma^2(3\cos^2\theta-1)}{r^4}+O(r^{-5}),\nonumber\\
\gamma&=&\frac{M}{2r^2}+\frac{(2\sqrt{2}-1)3iMa\cos\theta}{\sqrt{2}r^3}-\frac{3Ma^2(3\cos^2\theta-1)}{4r^4}
+O(r^{-5}),\nonumber\\
\nu&=&\frac{3iMa\sin\theta}{4\sqrt{2}r^3}-\frac{Ma^2\sin\theta\cos\theta}{\sqrt{2}r^4}+O(r^{-5}).
\end{eqnarray}

3) Weyl curvature
\begin{eqnarray}
\Psi_0&=&\frac{3Ma^2\sin^2\theta}{r^5}+O(r^{-6}),\nonumber\\
\Psi_1&=&\frac{3iMa\sin\theta}{\sqrt{2}r^4}-\frac{12Ma^2\sin\theta\cos\theta}{\sqrt{2}r^5}+O(r^{-6}),\nonumber\\
\Psi_2&=&-\frac{M}{r^3}-\frac{3iMa\cos\theta}{4r^4}+\frac{3Ma^2(3\cos^2\theta-1)}{r^5}+O(r^{-6}),\nonumber\\
\Psi_3&=&-\frac{3iMa\sin\theta}{2\sqrt{2}r^4}
+\left[-\frac{3i}{2\sqrt{2}}M^2a\sin\theta+\frac{6i}{\sqrt{2}}Ma^2\sin\theta\cos\theta\right]r^{-5}+O(r^{-6}),\nonumber\\
\Psi_4&=&\frac{3Ma^2\sin^2\theta}{4r^5}+O(r^{-6}).
\end{eqnarray}

\section*{Appendix B}
Some spin-weight harmonics
\begin{eqnarray}
Y_{0,0}&=&\frac{1}{\sqrt{4\pi}};\nonumber\\
{}_1Y_{1,1}&=&\sqrt{\frac{3}{16\pi}}(\cos\theta+1)e^{i\phi},\nonumber\\
{}_1Y_{1,0}&=&\sqrt{\frac{3}{8\pi}}\sin\theta,\nonumber\\
{}_1Y_{1,-1}&=&\sqrt{\frac{3}{16\pi}}(1-\cos\theta)e^{-i\phi};\nonumber\\
Y_{1,1}&=&-\sqrt{\frac{3}{8\pi}}\sin\theta e^{i\phi},\nonumber\\
Y_{1,0}&=&\sqrt{\frac{3}{4\pi}}\cos\theta,\nonumber\\
Y_{1,-1}&=&\sqrt{\frac{3}{8\pi}}\sin\theta e^{-i\phi};\nonumber\\
{}_{-1}Y_{1,1}&=&\sqrt{\frac{3}{16\pi}}(1-\cos\theta)e^{i\phi},\nonumber\\
{}_{-1}Y_{1,0}&=&-\sqrt{\frac{3}{8\pi}}\sin\theta,\nonumber\\
{}_{-1}Y_{1,-1}&=&\sqrt{\frac{3}{16\pi}}(1+\cos\theta)e^{-i\phi};\nonumber\\
{}_{2}Y_{2,2}&=&3\sqrt{\frac{5}{96\pi}}(1+\cos\theta)^2e^{2i\phi},\nonumber\\
{}_{2}Y_{2,1}&=&3\sqrt{\frac{5}{24\pi}}\sin\theta(1+\cos\theta)e^{i\phi},\nonumber\\
{}_{2}Y_{2,0}&=&3\sqrt{\frac{5}{16\pi}}\sin^2\theta,\nonumber\\
{}_{2}Y_{2,-1}&=&3\sqrt{\frac{5}{24\pi}}\sin\theta(1-\cos\theta)e^{-i\phi},\nonumber\\
{}_{2}Y_{2,-2}&=&3\sqrt{\frac{5}{96\pi}}(1-\cos\theta)^2e^{-2i\phi};\nonumber\\
{}_{1}Y_{2,2}&=&-3\sqrt{\frac{5}{24\pi}}\sin\theta(1+\cos\theta)e^{2i\phi},\nonumber\\
{}_{1}Y_{2,1}&=&3\sqrt{\frac{5}{24\pi}}(2\cos\theta-1)(1+\cos\theta)e^{i\phi},\nonumber\\
{}_{1}Y_{2,0}&=&3\sqrt{\frac{5}{4\pi}}\sin\theta\cos\theta,\nonumber\\
{}_{1}Y_{2,-1}&=&3\sqrt{\frac{5}{24\pi}}(2\cos\theta+1)(1-\cos\theta)e^{-i\phi},\nonumber\\
{}_{1}Y_{2,-2}&=&3\sqrt{\frac{5}{24\pi}}\sin\theta(1-\cos\theta)e^{-2i\phi};\nonumber\\
{}_{}Y_{2,2}&=&3\sqrt{\frac{5}{16\pi}}\sin^2\theta e^{2i\phi},\nonumber\\
{}_{}Y_{2,1}&=&-6\sqrt{\frac{5}{16\pi}}\sin\theta\cos\theta
e^{i\phi},\nonumber\\
{}_{}Y_{2,0}&=&\sqrt{\frac{5}{24\pi}}(3\cos^2\theta-1),\nonumber\\
{}_{}Y_{2,-1}&=&6\sqrt{\frac{5}{16\pi}}\sin\theta\cos\theta
e^{-i\phi},\nonumber\\
{}_{}Y_{2,-2}&=&3\sqrt{\frac{5}{16\pi}}\sin^2\theta e^{-2i\phi};\nonumber\\
{}_{-1}Y_{2,2}&=&-3\sqrt{\frac{5}{24\pi}}\sin\theta(1-\cos\theta)e^{2i\phi},\nonumber\\
{}_{-1}Y_{2,1}&=&3\sqrt{\frac{5}{24\pi}}(2\cos\theta+1)(1-\cos\theta)e^{i\phi},\nonumber\\
{}_{-1}Y_{2,0}&=&-\sqrt{\frac{5}{4\pi}}\sin\theta\cos\theta,\nonumber\\
{}_{-1}Y_{2,-1}&=&3\sqrt{\frac{5}{24\pi}}(2\cos\theta-1)(1+\cos\theta)e^{-i\phi},\nonumber\\
{}_{-1}Y_{2,-2}&=&3\sqrt{\frac{5}{24\pi}}\sin\theta(1+\cos\theta)e^{-2i\phi};\nonumber\\
{}_{-2}Y_{2,2}&=&3\sqrt{\frac{5}{96\pi}}(1-\cos\theta)^2e^{2i\phi},\nonumber\\
{}_{-2}Y_{2,1}&=&-3\sqrt{\frac{5}{24\pi}}\sin\theta(1-\cos\theta)e^{i\phi},\nonumber\\
{}_{-2}Y_{2,0}&=&3\sqrt{\frac{5}{16\pi}}\sin^2\theta,\nonumber\\
{}_{-2}Y_{2,-1}&=&-3\sqrt{\frac{5}{24\pi}}\sin\theta(1+\cos\theta)e^{-i\phi},\nonumber\\
{}_{-2}Y_{2,-2}&=&3\sqrt{\frac{5}{96\pi}}(1+\cos\theta)^2e^{-2i\phi}.\nonumber
\end{eqnarray}


\end{document}